\documentclass[global,draft]{svjour}
\usepackage{latexsym}
\usepackage{graphics,epsfig}
\usepackage{graphicx}
\journalname{General Relativity and Gravitation}

\begin{document}

\title{Brane Cosmology And Motion Of Test Particles In Five-Dimensional Warped Product Spacetimes}

\author{Sarbari Guha\inst{1}\and Subenoy Chakraborty\inst{2}}

\institute{Department of Physics, St. Xavier's College (Autonomous), Kolkata 700 016, India \\ \email {sarbariguha@rediffmail.com} \and Department of Mathematics, Jadavpur University, Kolkata 700032, India \\ \email{schakraborty@math.jdvu.ac.in}}

\date{Received: \today / Accepted: date / Published online: date \\ \copyright{ The Author 2009}}

\titlerunning{Brane Cosmology And Motion Of Test Particles}
\maketitle

\begin{abstract}
In the "braneworld scenario" ordinary standard model matter and non-gravitational fields are confined by some trapping mechanism to the 4-dimensional universe constituting the D3-branes which are embedded in a (4 + n)-dimensional manifold referred to as the 'bulk' (n being the number of extra dimensions). The notion of particle confinement is necessary for theories with non-compact extra dimensions, otherwise, the particles would escape from our 4-dimensional world along unseen directions. In this paper, we have considered a five-dimensional warped product space-time having an exponential warping function which depends both on time as well as on the extra coordinates and a non-compact fifth dimension. Assuming that the lapse function may either be a constant or a function of both time and of the extra coordinates, we have studied the nature of the geodesics of test particles and photons and have analyzed the conditions of stability in this geometrical framework. We have also discussed the possible cosmology of the corresponding (3 + 1)-dimensional hypersurfaces.
\end{abstract}


\section{Introduction}
\label{sec:intro}

According to the string theory postulate, standard model matter and non-gravitational fields are confined by some trapping mechanism to the 4-dimensional universe constituting the D3-branes (4-dimensional timelike hypersurfaces) that are embedded in a (4+n)-dimensional manifold referred to as the 'bulk' (n being the number of extra dimensions). This has led to a renewed interest in extra-dimensional theories of gravity. The success of the Kaluza-Klein \cite{kk} theory in particle physics led several workers to construct numerous models of non-compact higher-dimensional theories of physics \cite{Joseph}-\cite{WessPon}, and eventually to the so-called "braneworld scenario" \cite{add1}-\cite{add2}. In the braneworld models with non-compact extra dimension, particles and non-gravitational fields are assumed to be confined to the branes. At low energies, gravity is localized at the brane along with the particles but at high energies gravity "leaks" into the higher-dimensional bulk and is propagated therein. These "braneworld" models have been used to address several issues, one of which is to explain why the observable universe is found to be 4-dimensional \cite{rs1},\cite{rs2}. Consequently, it is necessary to determine how closely the corresponding apparent 4-dimensional world resemble the observed world \cite{anderson1}, and thus we need to verify the extent to which these theories satisfy the geodesic postulates concerning the motion of test particles and light rays.

The notion of particle confinement is necessary for theories with non-compact extra dimensions, as otherwise, the particles would escape from our 4-dimensional world along unseen directions. In the classical context, confinement of a test particle to the brane eliminates the effects of extra dimensions, thereby rendering them undetectable. In the braneworld scenario, the stability of the confinement of matter fields at the quantum level is made possible by assuming an interaction of matter with a scalar field. It appears that non-gravitational forces acting in the bulk and orthogonal to the spacetime are necessary to keep the test particles moving on the brane, the source of these confining forces being interpreted in different ways \cite{Rubakov}-\cite{dahia2}. The alternative explanation in case of higher-dimensional theories involve the invocation of geometrical mechanisms to support lower-dimensional confinement. In this case, confinement is due purely to the classical gravitational effects, without requiring the presence of brane-type confinement mechanisms. This has led to the investigation of warped product spaces and their geometrical properties \cite{felder}. Randall and Sundrum \cite{rs1},\cite{rs2}, achieved it through the construction of five-dimensional warped product spaces using an exponential warp factor in a non-factorizable metric furnished with mirror symmetry, even when the fifth dimension was infinite. In case of warped product spaces, it has been observed that a general qualitative analysis of the behavior of massive particles and photons in the fifth dimension can be made from the knowledge of the warping function. A general picture of these geodesic motions has been obtained by using the natural decoupling occurring between motions in the brane and motion in the fifth dimension in case of such spaces \cite{dahia3},\cite{dahia4}. For a thin asymmetric braneworld, bulk and brane geodesics does not coincide in general. However, if appropriate energy conditions are satisfied by the matter confined to the brane, then in presence of mirror-symmetry, test particles can be confined by gravity to a small region about the brane \cite{SW},\cite{Seahra}. The $Z_{2}$-symmetric braneworlds have their apparent and bulk geodesics coinciding on the brane. Although null bulk geodesic motion in the RS2-type
braneworlds, as well as in the static universe in the bulk of a charged topological AdS black hole have been studied \cite{Youm}, the same is not found for timelike geodesics.

Another important program in the study of higher dimensional models is the cosmological interpretation of the corresponding 4-dimensional geometry. By use of the Einstein equations, it has been shown that \cite{eisenhart} the embedding of a surface in a flat space of co-dimension one imposes the restriction that the surface has a constant curvature, if its dimension is $n > 2$. The effective equations for gravity in four dimensions were obtained by Shiromizu et al \cite{shiromizu} and subsequently cosmological solutions have been studied by some authors \cite{ida},\cite{mukho}.

In this paper, we have considered RS-type braneworlds with the bulk in the form of a five-dimensional warped product space-time, having an exponential warping function which depends both on time as well as on the extra coordinates and a non-compact fifth dimension. We know that the exponential warp factor reflects the confining role of the bulk cosmological constant \cite{lrr} to localize gravity at the brane through the curvature of the bulk. It is, therefore, possible that such localization may include some time-dependence and hence the choice. Employing the technique used by Dahia \cite{dahia3}, we have obtained a mechanism for the confinement of geodesics on co-dimension one hypersurfaces,
considering that the confinement is purely due to the classical gravitational effects. We have been able to obtain the description of the geodesic motions by using the natural splitting occurring between the motion in the extra dimension and the motion in the four-dimensional hypersurfaces. Such splitting helps us to use the phase space analysis to determine the nature of the geodesic motions in the neighbourhood of the hypersurfaces and analyze the conditions of stability. Further, we have assumed the lapse function to be either a constant or a function of both time and of the extra coordinates. This is possible since, in our case, the square of the lapse function is identical to the metric coefficient for the fifth dimension, which may therefore be of such a type. Finally, the cosmological interpretations of the corresponding (3 + 1)-dimensional hypersurfaces have also been discussed on the basis of the field equations.

\section{Geometric Construction of Five-Dimensional Warped Product Spaces}
\label{sec:1}

A warped product space \cite{BisNeill},\cite{CarCos}, is constructed as follows: Let us consider two manifolds (Riemannian or semi-Riemannian) $(M^{m}, h)$ and $(M^{n}, \bar{h})$ of dimensions $m$ and $n$, with metrics $h$ and $\bar{h}$ respectively. Given a smooth function $f : M^{n} \rightarrow \Re$ (henceforth called the 'warping function'), we can build a new Riemannian (or semi-Riemannian) manifold $( M , g )$ by setting $M = M^{m} \times M^{n}$, which is defined by the metric $g = e^{2f}h\bigoplus \bar{h}$. Here (M, g) is called a 'warped product manifold'. The case dim $M=4$ corresponds to (M, g) being a spacetime, and is called a 'warped product spacetime' (or simply 'warped spacetime').

In local coordinates ${z^{A}}$, the line element corresponding to the bulk metric is denoted by

\begin{equation}
dS^{2}= g_{AB}dz^{A}dz^{B}.
\end{equation}

The class of warped geometries considered by us is represented in general by the line element of the form
\begin{equation}
dS^{2}= e^{2f}h_{\alpha\beta}dx^{\alpha}dx^{\beta} + \bar{h}_{ab}dy^{a}dy^{b},
\end{equation}
where the extra coordinates are represented by $y^{a}$, the coordinates on the m-dimensional submanifold by $x^{\mu}$, $f$ is a scalar function, $h_{\alpha\beta}=h_{\alpha\beta}(x)$ is the warp metric on the submanifold of dimension \emph{m} and $\bar{h}_{ab}$ is the metric representing the extra dimensional part.

In this paper, we consider $m=4$ and $n=1$, so that $M = M^{4} \times M^{1}$, where $M^{4}$ is a Lorentz manifold with signature (+ - - -). The canonical metric due to Mashhoon et. al. \cite{MLW},\cite{WML} is more general than the Randall-Sundrum metric and encompasses all the metrics usually considered in the braneworld and Induced Matter Theory (IMT) approaches \cite{WessPon},\cite{IMT}-\cite{Wesson2}. For this, the physical metric in 4-dimension is assumed to be conformally related to the induced metric \cite{Ponce2}, i.e.
\begin{equation}
dS^{2}= \Omega ds^{2} + \varepsilon\Phi^{2}dy^{2},
\end{equation}
where $\Omega$ is the "warp" factor that satisfies the condition $\Omega>0$ and the scalar $\Phi$ which normalizes the vector normal to the hypersurfaces $y=constant$ is known as the \emph{lapse function}. We assume that the brane is defined by the $y=constant$ hypersurface, where $y$ is a Gaussian normal coordinate orthogonal to the brane, representing the fifth dimension, which is non-compact and curved (warped) \cite{lrr},\cite{rs2}.

In general, the warp factor is assumed to be a function of extra coordinates only, but here we assume that it is a function of both time, as well as of the extra coordinates. As in the RS models, gravity is localized on the brane through the curvature of the bulk. We know that the bulk cosmological constant acts to "squeeze" the gravitational field closer to the brane, with the exponential warp factor reflecting the confining role of the bulk cosmological constant \cite{lrr}. Mathematically, the time dependence of the warp factor does not affect its smooth nature. Physically, it takes into account the possibility that the confining role of the bulk cosmological constant and the curvature of the bulk may have some dependence on time, in addition to their dependence on the extra dimensional coordinate. When the extra dimension is spacelike, we have $\varepsilon=-1$. In our case, $\Phi^{2}=p=g_{yy}$ (which means that the square of the lapse function is identical to the metric coefficient for the fifth dimension) and $\Omega=e^{2f(t,y)}$. The expression for the line element therefore takes the form

\begin{equation}
dS^{2}= e^{2f(t,y)}h_{\alpha\beta}dx^{\alpha}dx^{\beta} - pdy^{2}.\label{e0}
\end{equation}

Here, p may either be a constant or a function of coordinates. We shall consider two cases, viz. $p=1$ and $p=p(t,y)$ (for example, the second type can be found in \cite{bdl}). To take into account the time-dependence of the warp factor, we assume the scalar function $f(t,y)$ to be of the form
\begin{equation}
f(t,y)=at+l\ln(\cosh(cy))\label{e00}
\end{equation}
where $a$, $l$  and $c$ are constants. We will not make any assumption on the topology of the extra dimension, so that, in principle $-\infty<y<+\infty$.

\section{Geodesic Motion}
\label{sec:2}

Dahia \cite{dahia3} demonstrated \emph{oscillatory confinement} of massive particles and light rays for branes of finite thickness in the context of 4 + 1-dimensional warped product spaces, whereby, the particles in the four-dimensional hypersurface can oscillate about the hypersurface, while remaining close to it. They have shown that such behaviour can occur for large classes of bulks which possess warped product geometries. Writing the geodesic equations for warped product spaces, they showed that the equation that describes the motion in 5D decouples from the rest. They made a qualitative analysis of the motions by rewriting the geodesic equation in the fifth dimension as an autonomous planar dynamical system and then employed the phase plane analysis to study the motion of particles with nonzero rest mass and photons respectively. Consequently they could draw very general conclusions about the possible existence of confined motions and their stability in the neighbourhood of the hypersurfaces. In our analysis, which now follows, we have used this method to arrive at the results.

The equations of geodesics in the 5-dimensional space M is given by
\begin{equation}
\frac{d^{2}z^{A}}{d\lambda^{2}}+^{(5)}\Gamma^{A}_{BC}\frac{dz^{B}}{d\lambda}\frac{dz^{C}}{d\lambda}=0, \label{e01}
\end{equation}
where $\lambda$ is an affine parameter and $ ^{(5)}\Gamma^{A}_{BC}$ are the 5-dimensional Christoffel symbols of the second kind defined by $ ^{(5)}\Gamma^{A}_{BC}=\frac{1}{2}g^{AD}(g_{DB,C}+g_{DC,B}-g_{BC,D})$. If we denote the fifth coordinate $z^{4}$ by $y$ and the remaining "spacetime" coordinates $z^{\mu}$ by $x^{\mu}$ i.e. $z^{A}=(x^{\mu},y)$, it can be shown that the 4-dimensional part of the geodesic equations (\ref{e01}) can be rewritten in the form \cite{dahia3}
\begin{equation}
\frac{d^{2}x^{\mu}}{d\lambda^{2}}+^{(4)}\Gamma^{\mu}_{\alpha \beta}\frac{dx^{\alpha}}{d\lambda}\frac{dx^{\beta}}{d\lambda}=\xi^{\mu}, \label{e02}
\end{equation}
where
\begin{equation}
\xi^{\mu}=-^{(5)}\Gamma^{\mu}_{44}\left(\frac{dy}{d\lambda}\right)^{2}-2^{(5)}\Gamma^{\mu}_{\alpha 4}\frac{dx^{\alpha}}{d\lambda}\frac{dy}{d\lambda}-\frac{1}{2}g^{\mu 4}(g_{4 \alpha,\beta}+g_{4 \beta,\alpha}-g_{\alpha \beta,4})\frac{dx^{\alpha}}{d\lambda}\frac{dx^{\beta}}{d\lambda}, \label{e03}
\end{equation}
with $^{(4)}\Gamma^{\mu}_{\alpha \beta}=\frac{1}{2}g^{\mu \nu}(g_{\nu \alpha,\beta}+g_{\nu \beta,\alpha}-g_{\alpha \beta,\nu})$.

\bigskip
We consider that the bulk spacetime is a 5-dimensional warped product spacetime in the braneworld scenario, represented by the line element (\ref{e0}) with the warping function given by (\ref{e00}). Let us now assume that this bulk space-time is foliated by a family of hypersurfaces defined by the equation $y=constant$. The geometry of each such leaves of foliation denoted by, say $y=y_{0}$ will be determined by the induced metric

\begin{equation}
ds^{2}=g_{\alpha\beta}(x,y_{0})dx^{\alpha}dx^{\beta}=e^{2f(t,y_{0})}h_{\alpha \beta}(x)dx^{\alpha}dx^{\beta}=\eta_{\alpha \beta}dx^{\alpha}dx^{\beta}.
\end{equation}

The extrinsic curvature of such a hypersurface is given by $2 K_{\alpha\beta}=-(\frac{\partial \eta_{\alpha \beta}}{\partial y})$. In that case, the quantities $^{(4)}\Gamma^{\mu}_{\alpha \beta}$ which appear on the left side of Eq. (\ref{e02}), are the Christoffel symbols associated with the induced metric in the leaves of foliation defined above.

\subsection{Case 1: p=1}

For the class of warped geometries given by Eq. (\ref{e0}) and (\ref{e00}), we can easily see that for $p=1$ we have $^{(5)}\Gamma^{\mu}_{44}=0$ and $^{(5)}\Gamma^{\mu}_{\alpha 4}=\frac{1}{2}g^{\mu \nu}\left(\frac{\partial g_{\nu \alpha}}{\partial y}\right)$.

In this case, the 4-dimensional part of the geodesic equations reduces to the form
\begin{equation}
\frac{d^{2}x^{\mu}}{d\lambda^{2}}+^{(4)}\Gamma^{\mu}_{\alpha \beta}\frac{dx^{\alpha}}{d\lambda}\frac{dx^{\beta}}{d\lambda}=-2f^{\prime}\frac{dx^{\mu}}{d\lambda}\frac{dy}{d\lambda}.\label{e04a}
\end{equation}
where a prime denotes differentiation with respect to $y$. Similarly, the geodesic equation for the fifth coordinate y in this warped product space becomes

\begin{equation}
\frac{d^{2}y}{d\lambda^{2}}+f^{\prime} e^{2f}h_{\alpha\beta}\frac{dx^{\alpha}}{d\lambda}\frac{dx^{\beta}}{d\lambda}=0.\label{e04}
\end{equation}

The above two geodesic equations (\ref{e04a}) and (\ref{e04}) are identical to those in Ref. \cite{dahia3}. This happens due to the special nature of the warping function chosen by us. We like to point out that the type of warping function in our case is different from that used in
\cite{dahia3}, as they considered a warping function which depends on the extra coordinates only. So, we can conclude that the geodesic equations do not depend on the explicit form of the warping function for this particular choice when $p=1$.  The result can be stated in the form of the following proposition:

\bigskip
\emph{\textbf{Proposition1}}: The geodesic equations for the 4-dimensional spacetime and for the fifth dimension remains the same as in the case with a warp factor which depends only on the extra coordinate, even after the inclusion of the additional linear time-dependence of the warping function, provided $g_{yy}=1$.

\bigskip
A study of the 5-dimensional motion of particles with non-zero rest mass near the hypersurface can be done by considering the 5-dimensional timelike geodesics $(g_{AB}\frac{dz^{A}}{d\lambda}\frac{dz^{B}}{d\lambda}=1)$, for which the above equation can be easily decoupled from the 4-dimensional spacetime coordinates as in \cite{dahia3}, yielding

\begin{equation}
\frac{d^{2}y}{d\lambda^{2}}+f^{\prime}\left(1+ \left(\frac{dy}{d\lambda}\right)^{2}\right)=0.\label{e05}
\end{equation}

The motion of photons in 5-dimension can similarly be analyzed by considering the null geodesics, for which $(g_{AB}\frac{dz^{A}}{d\lambda}\frac{dz^{B}}{d\lambda}=0)$ and we get from Eq. (\ref{e04})

\begin{equation}
\frac{d^{2}y}{d\lambda^{2}}+f^{\prime}\left(\frac{dy}{d\lambda}\right)^{2}=0.\label{e06}
\end{equation}

Equations (\ref{e05}) and (\ref{e06}) are second order ordinary differential equations, which can be solved if we know the nature of the warping function $f$.

\subsubsection{Motion of particles in four-dimensional spacetime}

To determine the nature of motion of particles in the four-dimensional world, we need to find the constraints on the warping function that will determine whether the geodesics on the bulk manifold may coincide with those on the hypersurface or not. Considering the expression for the extrinsic curvature of the hypersurface in this case, we find that the result is identical to those in Ref. \cite{dahia4}.

\subsubsection{Motion of particles in the fifth dimension}

A qualitative analysis of the motion in fifth dimension can be done without actually solving Eqs. (\ref{e05}) and (\ref{e06}) by defining $q=\frac{dy}{d\lambda}$ and investigating the autonomous dynamical systems \cite{WE}
\begin{equation}
\frac{dy}{d\lambda}=q \label{e07}
\end{equation}
\begin{equation}
\frac{dq}{d\lambda}=P(q,y) \label{e08}
\end{equation}
with $P(q,y)=-f^{\prime}(\epsilon+q^{2})$, where $\epsilon=1$ for timelike geodesics and $\epsilon=0$ for null geodesics. The \emph{equilibrium points} of the system of equations (\ref{e07}) and (\ref{e08}) are given by $\frac{dy}{d\lambda}=0$ and $\frac{dq}{d\lambda}=0$. Knowledge of these points along with their stability properties can provide a lot of information about the behavior allowed by this dynamical system. For the warping function considered by us, the analysis is \emph{very similar} to the one done by Dahia et. al. \cite{dahia3}.

Therefore, we find that the results obtained in  \cite{dahia3} and \cite{dahia4} are true even if we include a linear time-dependence of the warping function, provided $g_{yy}=1$.

\subsection{Case 2: p=p(t,y)}

The geodesic motion of the test particle in the bulk spacetime is still described by the equations (\ref{e01}). Let us consider a bulk metric of the form:
\begin{equation}
dS^2 =e^{2f(t,y)}\left(dt^2 -btdr^2 -btr^2d\theta^2 -btr^2sin(\theta)^2d\phi^2\right) -p(t,y)dy^2 \label{e31}
\end{equation}
where $b$ is another constant and $f$ is given by Eqn. (\ref{e00}) as $f(t,y)=at+l\ln(\cosh(cy))$. On substituting the explicit expression for the bulk metric, the geodesic equations reduce to the following five second order differential equations:

\begin{equation}
\frac{d^{2} t}{d \lambda^{2}} + a \left( \frac{dt}{d\lambda}\right)^{2}
+ \frac{(2at+1)b}{2}\left[ \left(\frac{dr}{d\lambda}\right)^{2} + r^{2}\left\{ \left( \frac{d\theta}{d\lambda}\right)^{2}
+ sin^{2}\theta \left( \frac{d\phi}{d\lambda}\right)^{2} \right\} \right] = 0, \label{e11}
\end{equation}

\begin{equation}
\frac{d^{2} r}{d \lambda^{2}}+ \frac{(2at+1)}{2t}\frac{dt}{d\lambda}\frac{dr}{d\lambda}
- r\left\{ \left( \frac{d\theta}{d\lambda}\right)^{2}+ sin^{2}\theta \left( \frac{d\phi}{d\lambda}\right)^{2} \right\}  = 0, \label{e12}
\end{equation}

\begin{equation}
\frac{d^{2} \theta}{d \lambda^{2}}+ \frac{d\theta}{d\lambda}\left\{\frac{(2at+1)}{2t}\frac{dt}{d\lambda}+\frac{1}{r}\frac{dr}{d\lambda}\right\}
-  sin\theta\cos\theta \left( \frac{d\phi}{d\lambda}\right)^{2}   = 0, \label{e13}
\end{equation}

\begin{equation}
\frac{d^{2} \phi}{d \lambda^{2}}+ \frac{d\phi}{d\lambda}\left\{\frac{(2at+1)}{2t}\frac{dt}{d\lambda}+\frac{1}{r}\frac{dr}{d\lambda}
+\frac{\cos\theta}{sin\theta}\frac{d\theta}{d\lambda}\right\}   = 0, \label{e14}
\end{equation}

\begin{equation}
\frac{d^{2}y}{d\lambda^{2}}+ \frac{1}{2p}\left\{2\frac{\partial p}{\partial t }\frac{dt}{d\lambda}\frac{dy}{d\lambda}
+ \frac{\partial p}{\partial y}\left(\frac{dy}{d\lambda} \right)^{2} \right\} = 0, \label{e15}
\end{equation}

Let us now analyse the equations of 5-dimensional motion of both photons as well as the particles of non-zero rest mass, near the hypersurface. For that we consider the equations of the 5-dimensional geodesics to be given by
\begin{equation}
g_{AB}\frac{dz^{A}}{d\lambda}\frac{dz^{B}}{d\lambda}=\epsilon, \label{e16}
\end{equation}
where, $\epsilon=1$ for timelike geodesics and $\epsilon=0$ for null geodesics. Using the expression for the bulk metric, we get from (\ref{e16}),

\begin{equation}
b\left[ \left(\frac{dr}{d\lambda}\right)^{2} + r^{2}\left\{ \left( \frac{d\theta}{d\lambda}\right)^{2}
+ sin^{2}\theta \left( \frac{d\phi}{d\lambda}\right)^{2} \right\} \right]\\
=\frac{1}{t}\left[\left( \frac{dt}{d\lambda}\right)^{2}-\frac{\left\{\epsilon+p\left(\frac{dy}{d\lambda}\right)^{2}\right\}}{e^{2f}}\right], \label{e17}
\end{equation}

Substituting from (\ref{e17}) in (\ref{e11}), we get

\begin{equation}
\frac{d^{2} t}{d \lambda^{2}} + a \left( \frac{dt}{d\lambda}\right)^{2}
+\frac{(2at+1)}{2t}\left[ \left(\frac{dt}{d\lambda}\right)^{2}-\frac{1}{e^{2f}}\left\{\epsilon+p\left(\frac{dy}{d\lambda}\right)^{2}\right\}\right]=0, \label{e18}
\end{equation}

Once again, a qualitative analysis of the motion of particles can be done without actually solving the Eqs. (\ref{e11}) to (\ref{e15}) by defining a dynamical system in the following way:

\begin{equation}
U=\frac{dt}{d\lambda}, \label{e19}
\end{equation}

and

\begin{equation}
V=\frac{dy}{d\lambda}, \label{e20}
\end{equation}

which, when substituted in (\ref{e18}) and (\ref{e15}) yields the equations

\begin{equation}
\frac{dU}{d\lambda}+\left[U^{2}\left(2a+\frac{1}{2t}\right)-\frac{1}{e^{2f}}\left\{\epsilon+pV^{2}\right\}\right]=0, \label{e21}
\end{equation}

and

\begin{equation}
\frac{dV}{d\lambda}+\frac{1}{2p}\left\{2UV\frac{\partial p}{\partial t}+V^{2}\frac{\partial p}{\partial y}\right\}=0. \label{e22}
\end{equation}

The set of equations (\ref{e19}) to (\ref{e22}) are used to analyse the evolution of the above dynamical system. The trajectories of this dynamical system can be described in a 2-dimensional phase space in terms of the variables $U$ and $V$. The system of equations (\ref{e19}) to (\ref{e22}) represent a real non-linear dynamical system \cite{R},\cite{LR} of the type
\begin{eqnarray}
\nonumber\frac{dt}{d\lambda} &=& U,\\
\nonumber\frac{dy}{d\lambda} &=& V,\\
\nonumber\frac{dU}{d\lambda} &=& P(U,V,t,y), \\
\frac{dV}{d\lambda} &=& Q(U,V,t,y), \label{e23}
\end{eqnarray}
where
\begin{eqnarray*}
 P(U,V,t,y) &=& -U^{2}\left(2a+\frac{1}{2t}\right)+\frac{1}{e^{2f}}\left\{\epsilon+pV^{2}\right\} \\
 Q(U,V,t,y) &=& -\frac{1}{2p}\left\{2UV\frac{\partial p}{\partial t}+V^{2}\frac{\partial p}{\partial y}\right\}
\end{eqnarray*}
with $\epsilon=1,0$ for timelike and null geodesics respectively. Both P and Q have continuous first partial derivatives for all (U,V). The evolution of the system in the (U,V) phase plane is given by the solution of the set (\ref{e23}). It is evident that the solution of the system for a given value of the parameter $\lambda$ is no longer unique, thereby, giving rise to very complicated trajectories. In such cases, the \emph{critical point} of the system must be a zero of the equations for all times for which the system is defined. Since no linear terms are present in the set (\ref{e23}), there will generally be several critical points of the system. We know that the zeros of the system correspond to the fixed points of the phase trajectories. For a nonautonomous system, generally each point of the phase space is intersected by many distinct trajectories. Therefore the collection of phase trajectories as well as the analysis for the identification of the fixed points becomes extremely complicated \cite{FM}. The dynamics of the phase trajectories is then strongly controlled by the underlying topology of the manifold under
consideration. Such an analysis is beyond the scope of the present paper. One simple method is to make suitable transformations to reduce the
non-autonomous systems to autonomous ones by appending $t$ to the depending variables. However, such a transformation has the drawback that the non-autonomous problem frequently looses its original algebraic structure. Hence it is not considered here.

\section{Phase trajectories and Critical Points of the system for the case p=p(t,y)}
\label{sec:3}
\bigskip
In spite of the fact that we are dealing with a nonautonomous system, we can still interpret the nature of the trajectories with the help of a simple analysis. The equation
\begin{equation}
\frac{dV}{dU}=\frac{Q}{P},\label{e25}
\end{equation}
at some given value of $y$ specifies the phase trajectory of the system in the (U,V) phase plane, provided $P\neq0$ at this point. Generally $P\neq0$ except at the point $U=V=0$ for null geodesics (which, as we shall find later, represents the critical point of the system in such a case). Substituting the explicit forms of P and Q in (\ref{e25}) and integrating, we arrive at the result
\begin{equation}
U^{2}V\left[\frac{1}{p}\frac{\partial p}{\partial t}-(2a+\frac{1}{2t})\right]+ UV^{2}\frac{1}{2p}\frac{\partial p}{\partial y} + V^3\frac{p}{3e^{2f}} +V\frac{\epsilon}{e^{2f}} + K =0,\label{e26}
\end{equation}
where K is the arbitrary constant of integration. Setting this constant to zero and simplifying, we can recast eqn. (\ref{e26}) into the form
\begin{equation}
AU^2 + BUV + CV^2 + D=0,\label{e26a}
\end{equation}
where $A$, $B$, $C$, and $D$ are functions of coordinates, specifically, $t$ and $y$. The above equation is useful to understand the phase trajectories even when no fixed points are found.

For \emph{null geodesics}, $D=0$ and (\ref{e26a}) reduces to
\begin{equation}
AU^2 + BUV + CV^2=0.\label{e26b}
\end{equation}
For a given value of $A$, $B$ and $C$, the phase trajectories are represented by a pair of straight lines intersecting at (0,0). The slope of the lines change for different choices of $A$, $B$ and $C$. It is evident that for null geodesics, $P=0=Q$ at $U=V=0$, which therefore represents the fixed point of the trajectories in this case.

However, for \emph{timelike geodesics}, $P\neq0$ even when $U=V=0$. In that case, eqn. (\ref{e26a}) represents the phase trajectory of the system even if no fixed point is found. In fact, we shall find from our subsequent analysis, that we cannot locate any fixed point for timelike geodesics, except for a special case. In the following figures, we have illustrated a few possible trajectories for some arbitrarily chosen values of the functions $A$, $B$, $C$, and $D$ in case of timelike geodesics. The Fig. 1 represents the plot for the sample equation  $U^2-6UV+8V^2=2$, Fig. 2 represents the equation $-U^2-6UV+8V^2=-2$, Fig. 3  represents the equation $U^2-6UV-8V^2=-2.5$ and Fig. 4 represents the equation $4U^2-6UV+8V^2=100$. We can interpret the different figures to correspond to different values of $y$ and hence they lie on different (U,V) phase planes. A particular trajectory lies on a given (U,V) phase plane with $y=constant$.

\begin{figure}[h]
\epsfig{file=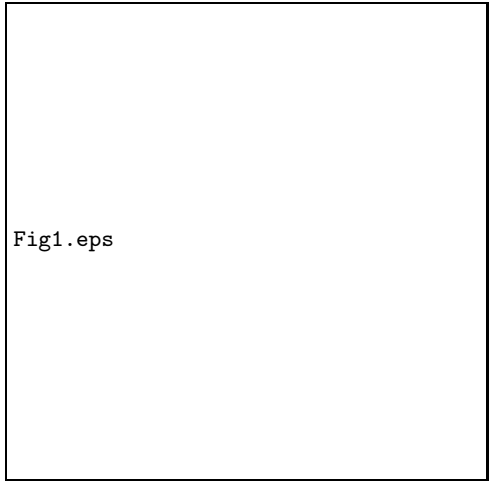,height=2.5in,width=2.5in}
\caption{Plot for the sample equation $U^2-6UV+8V^2=2$ for timelike geodesics. The trajectory appears to avoid the point (0,0).}
\label{fig:1}
\end{figure}

\begin{figure}[h]
\epsfig{file=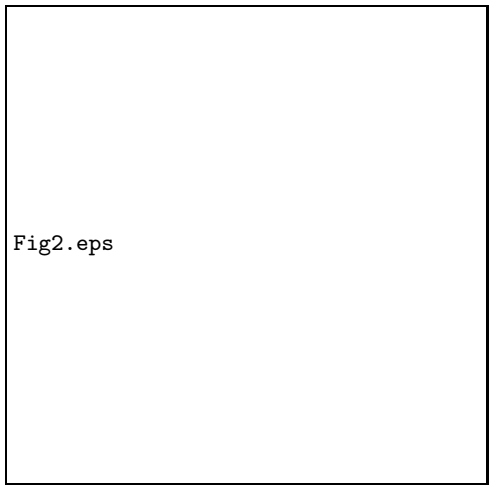,height=2.5in,width=2.5in}
\caption{Plot for the sample equation $-U^2-6UV+8V^2=-2$ for timelike geodesics. This trajectory also appears to move away from (0,0).}
\label{fig:2}
\end{figure}

\begin{figure}[h]
\epsfig{file=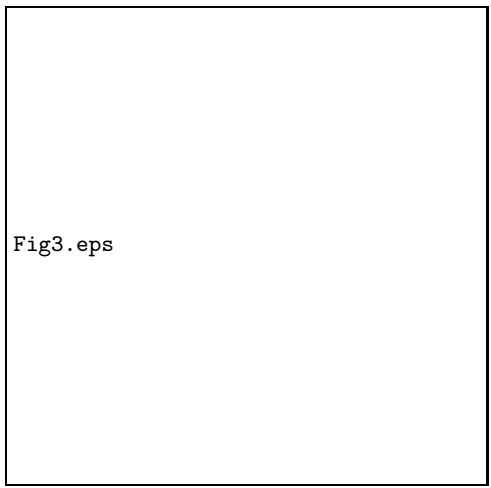,height=2.5in,width=2.5in}
\caption{Plot for the sample equation $U^2-6UV-8V^2=-2.5$ for timelike geodesics. This trajectory also behaves like the earlier ones.}
\label{fig:3}
\end{figure}

\begin{figure}[h]
\epsfig{file=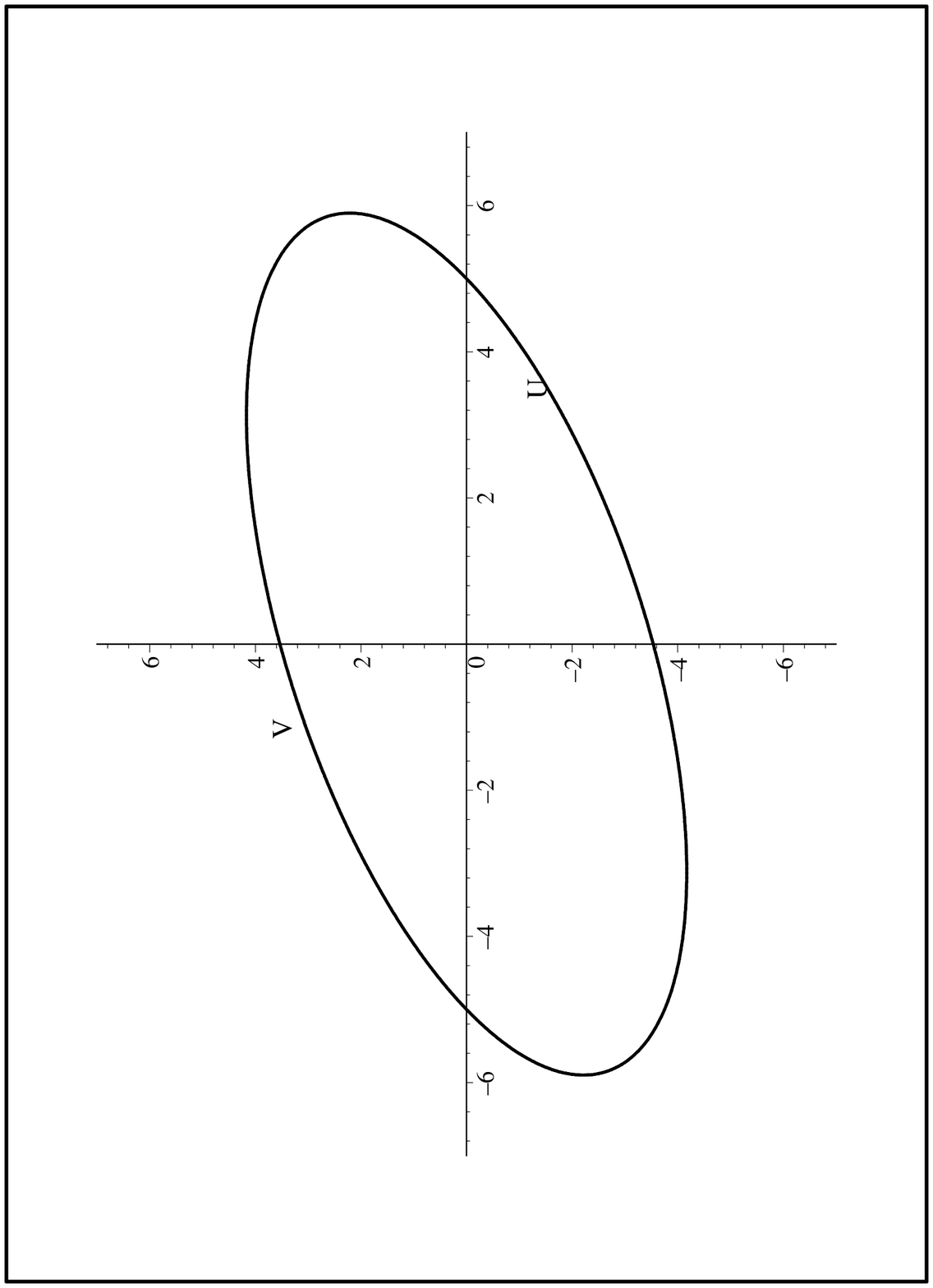,height=2.5in,width=2.5in}
\caption{Plot for the sample equation $4U^2-6UV+8V^2=100$ for timelike geodesics. This is a special case indicating a closed orbit, that signifies \emph{neutral equilibrium}.}
\label{fig:4}
\end{figure}

Initial examination of the above figures, indicates a few possibilities. Namely, the elliptic trajectory represents a closed orbit with $U=V=0$ as the \emph{center} and thus an example of \emph{neutral equilibrium}. Such an orbit represents a system whose total energy is conserved. In general, the trajectories appear to avoid the point $U=V=0$.

\bigskip
Thus when the specific warp factor is coupled with a lapse function, which depends on both time and extra coordinates, the
physical system is modeled by a nonautonomous dynamical system. This is different from the results obtained for $p=1$, as well as from those obtained by Dahia (\cite{dahia3} and \cite{dahia4}), who dealt with an autonomous system. We can therefore say that, the physical system in this case is a more general one.

\bigskip
We now proceed to examine the critical points of the system analytically. Let $(U_{0},V_{0})$ represent the critical point.

\bigskip
\emph{Null geodesics}: In this case, the two conditions $P=0$ and $Q=0$ lead us to

\begin{equation}
-U_{0}^{2}\left(2a+\frac{1}{2t}\right)+\frac{pV_{0}^{2}}{e^{2f}}=0,\label{e27}
\end{equation}

and

\begin{equation}
2U_{0}V_{0}\frac{\partial p}{\partial t}+V_{0}^{2}\frac{\partial p}{\partial y}=0.\label{e28}
\end{equation}

From (\ref{e27}), we have

\begin{equation}
V_{0}=U_{0}\sqrt{\frac{e^{2f}}{p}\left(2a+\frac{1}{2t}\right)}.\label{e29}
\end{equation}

Substituting from (\ref{e29}) in (\ref{e28}), we get
\begin{equation}
U_{0}^{2}\left[2W\frac{\partial p}{\partial t}+W^2\frac{\partial p}{\partial y} \right]=0,\label{e30}
\end{equation}
where, $W=\sqrt{\frac{e^{2f}}{p}\left(2a+\frac{1}{2t}\right)}$ and is not zero in general. Thus, in general, (\ref{e30}) will be valid if we have $U_{0}=0$ and hence, $V_{0}=0$. Thus (0,0) will be the critical point for null geodesics.
\bigskip

\emph{Timelike geodesics}: In this case, the condition $P=0=Q$ lead us to the result

\begin{equation}
U_{0}^{2}=\frac{\frac{1}{p}\left(\frac{\partial p}{\partial y}\right)^2}{W^2\left(\frac{\partial p}{\partial y}\right)^2-4\left(\frac{\partial p}{\partial t} \right)^2 },
\end{equation}

and

\begin{equation}
V_{0}^{2}=\frac{4\left(\frac{\partial p}{\partial t} \right)^2}{p\left\{W^2\left(\frac{\partial p}{\partial y}\right)^2-4\left(\frac{\partial p}{\partial t} \right)^2 \right\}}.
\end{equation}

Owing to our assumption regarding the nature of p, neither of $U_{0}$ or $V_{0}$ can be zero in general, nor can be imaginary unless

\bigskip
$W^2\left(\frac{\partial p}{\partial y}\right)^2 <4\left(\frac{\partial p}{\partial t} \right)^2 $.

\bigskip
Thus, the critical point cannot be determined unless the exact form of the function $p(t,y)$ is known. In fact, there will not be any fixed point unless the condition $P=0=Q$ is maintained for all times for which the dynamical system is defined.

\section{Cosmological Interpretations}
\label{sec:4}

The cosmology of the 4-dimensional hypersurfaces is determined by computing the Einstein tensor directly from the metric corresponding to the 5-dimensional spacetime. We have used the GRTensor package \cite{lake} for our calculations. For the type of line element given by (\ref{e0}) and (\ref{e00}), we choose the particular case,

\begin{equation}
dS^2 =e^{(2at+2l\ln(cosh(cy)))}\left(dt^2 -btdr^2 -btr^2d\theta^2 -btr^2sin(\theta)^2d\phi^2\right) -dy^2
\end{equation}

The 5-dimensional metric in general represents an observable universe with isotropic pressure given by $p_{r}=p_{\theta}=p_{\phi}=\frac{1}{4}\left(\frac{8at+4a^2t^2-1}{e^{(2at+2l\ln(cosh(cy)))}t^2}\right)$.
Both ordinary matter in the bulk as well as the induced matter on the hypersurface exhibit isotropic pressure given by the above expression. The geometric properties of the bulk spacetime remains unchanged for all values of $c$ including zero.

\bigskip
\emph{\textbf{Proposition2}}: It is possible to have matter residing on the corresponding 4-dimensional hypersurface in the form of pure radiation. The geometric properties of the bulk spacetime remains unchanged for all values of $c$ including zero.

\bigskip
There are two important points to note. First, for $a=0$, the warping function reduces to the type given by Gremm \cite{Gremm}. In that case we have $G_{ab}=R_{ab}$, and the Ricciscalar is identically zero. The 5-dimensional Ricci tensor is then identical to the Ricci tensor of the corresponding 4-metric and the matter on the hypersurface is in the form of pure radiation with $G^{t}_{t}=\frac{3}{4t^2}$ and $G^{r}_{r}=G^{\theta}_{\theta}=G^{\phi}_{\phi}=-\frac{1}{4t^2}$. Secondly, the Einstein tensor for the 5-metric given by (\ref{e31}) is identical to the 5-metric obtained by the transformation $\ln(cosh(cy))\rightarrow 1$ and hence the kind of 4-dimensional geometry observed here can also be embedded in other 5-metrics obtained by such transformations.

It is interesting to note that the observable universe still represents matter with isotropic pressure even in the case where the lapse function is a function of both time and extra coordinates.

\section{Conclusions}
\label{sec:5}

The consideration of the confinement of the geodesics of particles and the cosmology associated with the corresponding 4-dimensional hypersurfaces is an important program in the braneworld scenario. In this paper we have done it for RS-type braneworlds with the bulk in the form of a five-dimensional warped product space-time. The warping function is an exponential function of both time and extra coordinates and the lapse function is either a constant or a function of both time and extra coordinates. In such a case, we have obtained a mechanism for the confinement of geodesics on co-dimension one hypersurfaces.

Considering that confinement is purely due to classical gravitational effects and geometric in nature, we have obtained a description of the geodesic motions for these warped product spaces by using the natural splitting that occurs between the motion in the extra dimension and the motion in the four-dimensional hypersurfaces. Further, this splitting helps us to use the phase space analysis to determine the nature of the geodesic motions in the neighbourhood of hypersurfaces. It is observed that the geodesic equations for the 4-dimensional spacetime and for the fifth dimension do not depend on the time dependence of the warping function provided the lapse function is a constant. Further, in the general case when the lapse function is a function of both time and extra coordinates, using the dynamical system analysis we have determined the nature of the trajectories and have examined the critical points on a given hypersurface.

From the point of view of cosmology, the observable universe in general represents matter with isotropic pressure and which, in a particular case will be in the form of pure radiation.

For future work it is interesting to study whether the time dependence of the warp factor is related to the time evolution of the dark energy.

\section{Acknowledgments}
\label{sec:6}

The authors are thankful to the reviewers for their valuable comments and suggestions. SG thanks the University Grants Commission of the Government of India for financial support and to IUCAA, INDIA for an associateship.

\end{document}